\documentclass[12pt,preprint]{aastex}
\usepackage{amsmath}
\usepackage[normalem]{ulem} 

\def\bea{\begin{eqnarray}}
\def\eea{\end{eqnarray}}
\def\be{\begin{equation}}
\def\ee{\end{equation}}

\newcommand{\showlabel}[1]{} 
\newcommand{\internalcom}[1]{}

\received{}
\accepted{}
\journalid{}{}
\articleid{}{}
\paperid{}
\cpright{AAS}{2011}
\ccc{}

\shorttitle{Nucleation of Carbon-rich Structures}
\shortauthors{Patra, Kr\'al, and Sadeghpour}

\begin{document}

\title{ 
Nucleation and stabilization of carbon-rich structures in interstellar
media}

\author{
N. Patra\altaffilmark{1,3}, 
P. Kr\'al \altaffilmark{1,2},
and H. R. Sadeghpour\altaffilmark{3}} 

\altaffiltext{1}{
Department of Chemistry, University of Illinois at
Chicago, Chicago, IL 60607, USA; npatra2@uic.edu}

\altaffiltext{2}{ Department of Physics, University of Illinois at
Chicago, Chicago, IL 60607, USA; pkral@uic.edu}

\altaffiltext{3}{ITAMP, Harvard-Smithsonian Center for Astrophysics, 
Cambridge, MA 02138, USA; hrs@cfa.harvard.edu}

\begin{abstract}
We study conditions under which carbon clusters of different sizes 
form and stabilize. {We describe an approach to equilibrium by simulating tenuous carbon gas dynamics to long times.} First, we use
reactive molecular dynamics simulations to describe the nucleation of long
chains, large clusters, and complex cage structures in carbon and hydrogen
rich interstellar gas phases. We study how temperature, particle density,
presence of hydrogen, and carbon inflow affect the nucleation of molecular
moieties with different characteristics, in accordance with astrophysical
conditions. We extend the simulations to densities which are orders of
magnitude lower than current laboratory densities, to temperatures
relevant to circumstellar environments of planetary nebulae, and to longtime
(microsecond) formation timescales. We correlate cluster size distributions
from dynamical simulations with thermodynamic equilibrium intuitions, 
where at low temperatures and gas densities, entropy plays a significant role.
\end{abstract}


\section{Introduction}

Carbon-based chemistry of the interstellar medium (ISM) is now an established
discipline. Presence of large variety of organic molecules such as 
aldehydes \cite{marcelino07}, alcohols \cite{friberg88,marcelino07}, 
ethers \cite{herbst06}, amides \cite{hollis06}, sugar \cite{hollis00}, 
acids \cite{herbst06,gibb04}, amino acids \cite{botta02,cronin93}, and 
nitriles \cite{belloche08}, as well as many long-chain hydrocarbon compounds in
the dense interstellar medium have been confirmed by infrared, radio,
millimeter, and submillimeter spectroscopic measurement
\cite{Irvine1987,Ohishi1998, Winnewisser1999,Schutte1996,herbst09,hollis00}.
Detection of ring species, including aromatic rings, raises the possibility
that biologically significant molecules may exist in the ISM. Between 10\%-30\%
of the interstellar carbon budget is thought to reside in polycyclic aromatic
hydrocarbons (PAHs) \cite{tielens}. 

The presence of complex molecules, with typical spectral properties, can
be used to precisely probe the environments in which they reside and
provide a fairly faithful history of ISM molecular clouds. IR emission
lines (in the range of 3-14 $\mu$m) have been assigned to C-C and C-H
vibration bands of a broad class of polycyclic aromatic hydrocarbons
(PAHs) \cite{salama08, Ehrenfreund, Duley, Salama, Ricks, herbst09,
tielens}.  The simplest PAH, naphthalene (C$_{10}$H$_8$) has been found in
carbonaceous chondrites \cite{zare08}, suggesting extraterrestrial origin.
Although no specific PAH molecule has yet been unambiguously identified in
space, the infrared emission bands have been observed in nearly all
astronomical objects, from highly ionized HII regions to ultraluminous
infrared galaxies.  Moreover, the simplest PAH, naphthalene
(C$_{10}$H$_8$) has been already found in carbonaceous chondrites
\cite{zare08}, suggesting its extraterrestrial origin.  

The emission features at $7.0$, $8.5$, $17.4$ and $18.9$ $\mu$m, present in the
spectra of Tc1, NGC 7023 and NGC 2023 planetary and reflection nebulae, have
been conclusively associated with the vibrational lines of regular carbon cages
(C$_{60}$ and C$_{70}$) \cite{cami10,sellgren10}.  Moreover, a large class of
unsolved mysteries in astronomy, such as diffuse interstellar bands (DIB)
\cite{allamandolla85}, extended red emission (ERE) \cite{witt84}, and the
$2,175$ {\AA} ``bump" \cite{2175} trace their origins to large carbon and/or
carbonaceous molecules. Single and multi-layer graphene have strong plasmon
excitation bumps in the vicinity of the $2,175$ {\AA} feature
\cite{eberlein08}. However, more work in needed to establish potential
connection between these spectral features in astrophysical conditions. These
challenges can not be solved without a strong theoretical input. Development of
a comprehensive and detailed theoretical models for the origin and evolution of
complex organic molecules such as PAHs, fullerenes, and other ordered carbon
structures may be crucially important for understanding of the role that
prebiotic molecules played on early Earth.

Most theoretical studies of formation of larger aromatic carbon chains, rings,
and cages in ISM have been done close to equilibrium and at relatively high
particle densities. Naphthalene (C$_{10}$H$_8$) was shown to be
potentially formed in ISM via a barrierless reaction of phenyl radical
(C$_6$H$_5$) and vinylacetylene (C$_4$H$_4$) \cite{parker12}. It was
proposed that C$_{60}$ in ISM can form in a top-down manner in successive
de-hydrogenation of PAHs into graphene \cite{berne12}, and formation of defects
(pentagons) necessary for the synthesis of C$_{60}$. There are a number of
issues associated with this hypothesis of C$_{60}$ formation: 1) the
circumstellar environment of Tc1 planetary nebula is distinctly hydrogen poor,
2) graphene is only synthesized in the laboratory and is not believed to exist
in ISM, and 3) the proposed formation of carbon cages in ISM (via the 6 to 5
member rings) was observed in the laboratory only at far from equilibrium
conditions. 

To our knowledge, there are no available dynamical simulations of formation
of molecules with aromatic rings at astrophysically-relevant conditions,
which are not even properly known. Orderly growth of planar carbon
structures (hexagon rings) from precursor C$_2$ molecules necessitates
low-temperature conditions, where vibrational distortions of the structures
during their growth are minimized \cite{irle03}.  Under such conditions no
significantly curved (cage-like) structures are supposed to form \cite{irle03}.
Fast inflow of carbon material is needed to form curved carbon clusters and
fullerenes (pentagon rings, ring-stacks, ring fusion). In the laboratory,
fullerenes are created with remarkable efficiency ($\approx 40$ \% in a
condensing carbon arc) under less understood far from equilibrium
conditions. 

{In this work, we describe an approach to equilibrium by simulating 
tenuous carbon (and hydrogen) gas dynamics to exceeding long times. 
Different carbon molecular clusters and moieties which form represent 
steady state local equilibria and are important yardsticks for understanding 
future time dynamics and equilibration in astrophysical time.}

\vspace{-2mm}
\section{Simulations of cluster growth}
\vspace{-2mm}

We would like to understand the nucleation and stabilization of carbon
chains, rings, and cages of different sizes and moieties in regimes
relevant to astrophysics (cold temperatures and low densities), which are
orders of magnitude more dilute than those considered under laboratory
conditions \cite{irle03,irle06}. However, performing such simulations in
gaseous conditions, resembling an ISM, is a highly challenging task. Since
the cluster formation times are long and the systems are large, we cannot
reasonably investigate them by first principle molecular dynamics methods,
unless we focus only on equilibrium structures; for example, fullerene
forms by laser irradiation or arc-discharge methods within one ms to
several seconds.  By increasing the density of carbon atoms, we can
effectively compress the nucleation time. However, this time compression
enhances the growth and dissociation of the formed clusters due to
collisions. Moreover, it dramatically alters the entropic contribution to
the cluster Gibbs energy and thus influences the cluster form and
stability.

Here, we use semiclassical reactive molecular dynamics (MD) simulations to
model the formation of clusters starting from an elementary carbon (and
hydrogen) gas.  To proceed with the simulations, we need to know the
reactive force fields, characterized by appropriate potential functions
describing the change of atom hybridization over time.  We use an
approximate description of the chemical bonds, and propose analytical fit
formulae for the structure formation times.

We apply the adaptive intermolecular reactive empirical bond-order
(AIREBO) potential function proposed by Stuart \cite{Stuart}, based on the
Brenner bond-order potential \cite{Brenner1}. This potential can be used
for chemical reactions and intermolecular interactions in condensed-phase
hydrocarbon systems such as liquids, graphite, and polymers.  The
potential function is given by 
\begin{eqnarray} E_{Total} & = & E_{REBO} + E_{LJ} + E_{tors} \, ,
\nonumber \\ E_{REBO} & = & V_{ij}^{R}(r_{ij}) + b_{ij}V_{ij}^{A}(r_{ij})
\, , 
\label{AIREBO} \end{eqnarray} 
where $V_{ij}^{R}$ and $V_{ij}^{A}$ are repulsive and attractive pairwise
potentials between atoms $i$ and $j$, separated by $r_{ij}$, and $b_{ij}$
is the bond-order term in a Tersoff-type potential \cite{Stuart}. The
dispersion and intermolecular interaction ($E_{LJ}$) are modeled with a
Lennard-Jones (LJ) 12-6 potential having a switching function
and a connectivity switch \cite{Stuart}. The torsional potential
($E_{tors}$) is proportional to bond weights that contribute to the
dihedral angle \cite{Stuart}. 

We model the self-assembly of clusters from C and/or H atom gases using
atomistic reactive MD simulations with the AIREBO potential \cite{Stuart}
as implemented in the LAMMPS package \cite{LAMMPS}. The systems are
modeled in the NVT ensemble with periodic boundary conditions applied. The
Langevin damping method \cite{Gaspard} with a damping coefficient of
$0.01$ ps$^{-1}$ is used to thermalize the systems;  the time step is
$0.5$ fs. Visualization and analysis of the trajectories are done by VMD
\cite{VMD}. 

The simulations are progressively more computational expensive with
increasing temperatures, as finer time steps are required (scaling is
linear with temperature). For instance, it takes $24$ hours to simulate
$10,000$ C atoms at $T=3,000$ K for a $t\approx 1$ ns simulation time on a
$120$-core machine with 2.6 GHz AMD Opteron processors. For the same
reason higher temperature ($T=2,000$ K) systems are simulated for upto
$t=50$ ns.

\vspace{-2mm}
\section{Results and discussion} 
\vspace{-2mm}

Evidence from theoretical studies suggests that high temperatures are
required to transform a flat structure to a curved moiety \cite{Popov}. We
examine in our simulations the type of structures that are formed at
different temperatures and gas densities.

\vspace{-2mm} 
\subsection{Nucleation of carbon clusters at different temperatures
and densities}
\vspace{-2mm} 

First, we simulate the nucleation of carbon clusters at different
temperatures ($T=300-3,000$ K). We place $512$ atoms in a box of $50
\times 50 \times 50$ nm$^3$, giving a carbon particle density of
$4.1\times 10^{-6}$ \AA$^{-3}$.  In the MD simulations, the system energy
is first minimized for a short time ($5$ ps). { In most implementations 
of MD algorithm, applicable to high density gases and biological 
complexes, the system is initially relaxed to a minimum energy, to avoid 
unnecessary coordinate overlaps or collisions. While in our simulations, 
due to low densities, this initial minimization is not necessary, we 
maintain it for a very short period.} Then, the systems are
heated to the target temperature and simulated for up to
$3$ $\mu$s.  

\begin{figure} 
\plotone{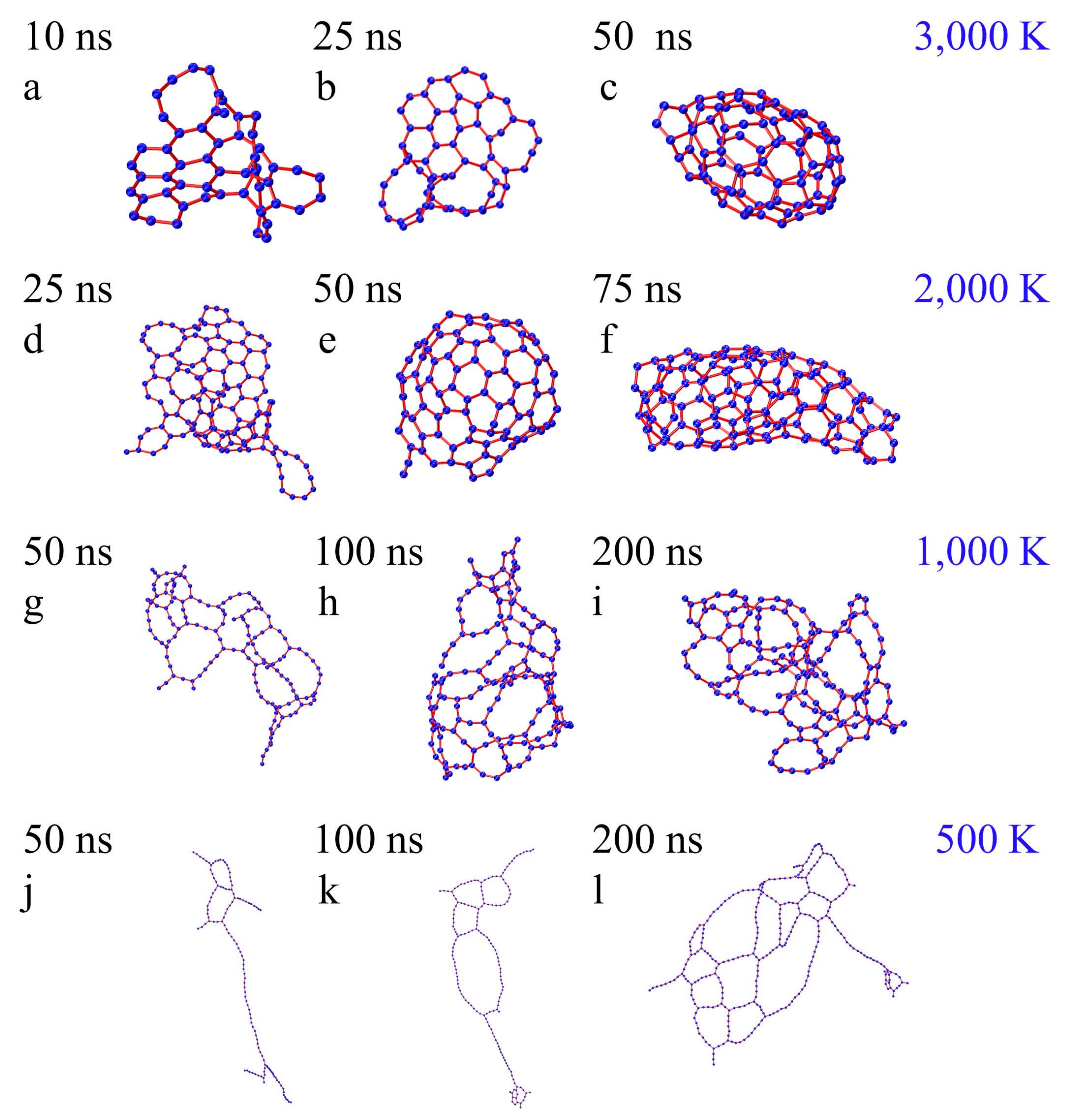}
\caption{Nucleation of carbon clusters at different
temperatures, $T$. (a--c: $T=3,000$ K) (typical structures are shown).  a)
Formation of planar clusters (six member rings) is observed within $10$
ns. b) Fullerene-type and big planar clusters with six and five member
rings are observed after $t=25$ ns.  c) Almost all carbon atoms form
fullerene-type clusters ($C_{70}-C_{84}$) within $t=50$ ns.  (d--e: $T=2,000$
K) d) Planar structures are formed within $t=25$ ns. e) Formation of
cage-like structures with five and six member rings are observed after
$t=50$ ns. f) Cylindrical clusters, due to the merger of small
fullerene-type clusters, are found within $t=75$ ns. (g--i: $T=1,000$ K) g)
Long chains emerge from short chains and eventually large member rings
structures are observed (snapshot taken at $t=50$ ns). h) Several large
member rings form cage-like structures (snapshot taken at $t=100$ ns). i)
Formation of five and six member rings are observed after $t=200$ ns.
(j--l: $T=500$ K) j) Long chains molecules are observed within $t=50$ ns. k)
Large member rings emerge from long chains after $t=100$ ns.  l) Planar
clusters with big rings are found after $t=200$ ns.  } \label{T}
\end{figure}

In Fig.~\ref{T}, we present snapshots of the carbon moieties formed at
different temperatures and times. At $T=3,000$ K, clusters appear at much
earlier times as the diffusion and collision rates are faster.  The high
thermal energy also helps to induce curvature in the flat structures at
much earlier times. As shown in Fig.~\ref{T}(top panel), planar clusters
(a) form within $t=10$ ns, and large planar clusters with five and six
member rings emerge after $t=25$ ns (b), while fully formed fullerene
structures ($C_{70}-C_{84}$) appear after $t=50$ ns (c). Even at early
times, defects (pentagons) are present in the moieties heralding the
process of curvature formation. After $t=90$ ns, nearly all structures
are cages. Although in our simulations (small systems), all carbon atoms
form curved structures, for a much bigger system at the same density and
at high temperature, predominantly smaller carbon clusters along with few
planar and curved moieties form when the system is in or close to
equilibrium (see distribution of cluster sizes section).

At $T=2,000$ K, the temperature is still high enough to overcome the
transition barriers (from flat structures to curved structures).
Snapshots are given at $t=25, 50$, and $75$ ns (d-f) of Fig.~\ref{T}. The
existence of five-member rings at $t=25$ ns (d) is a prelude to cage-like
structure formation. At $t=50$ ns (e), curvature is dramatic and a
fullerene-like structure begins to form. At $t=75$ ns, cylindrical
nano-structures emerge due to merger of fullerene-type clusters.  Again,
at $T=2,000$ K, the high thermal energy hinders the formation of large
clusters and small clusters are more likely to be present in equilibrium
(see distribution of cluster sizes section).

As we decrease the temperature, large clusters can be stabilized, as
shown in Fig.~\ref{T}(g-i). At $T=1,000$ K, even though defects are
forming at short times, $t=50$ ns (g), the tendency is to form chain-like
structures, while cage-like structure begin to appear after $t=100$ ns
(h). At $t=200$ ns (i), curvature is clearly evident, while regularity of
the type present at higher temperatures, is lacking. 

Long chains can be stabilized at lower temperatures. Snapshots at $t=50,
100$, and $200$ ns (j-l) in the lower panel of Fig.~\ref{T}, illustrate
that chain formation is readily accomplished at $T =500$ K with instances
of several irregularly formed rings. At $T=500$ K, cage like formation
takes very long times, if at all. At $t=200$ ns (l), there's evidence for
curvature formation, but rings with odd-number of atoms are quite
irregular in shape. For a much bigger system, carbon atoms can form
several large clusters and then they can form a ``liquid" phase at a
lower temperatures (see distribution of cluster sizes section). 

\begin{figure} 
\plotone{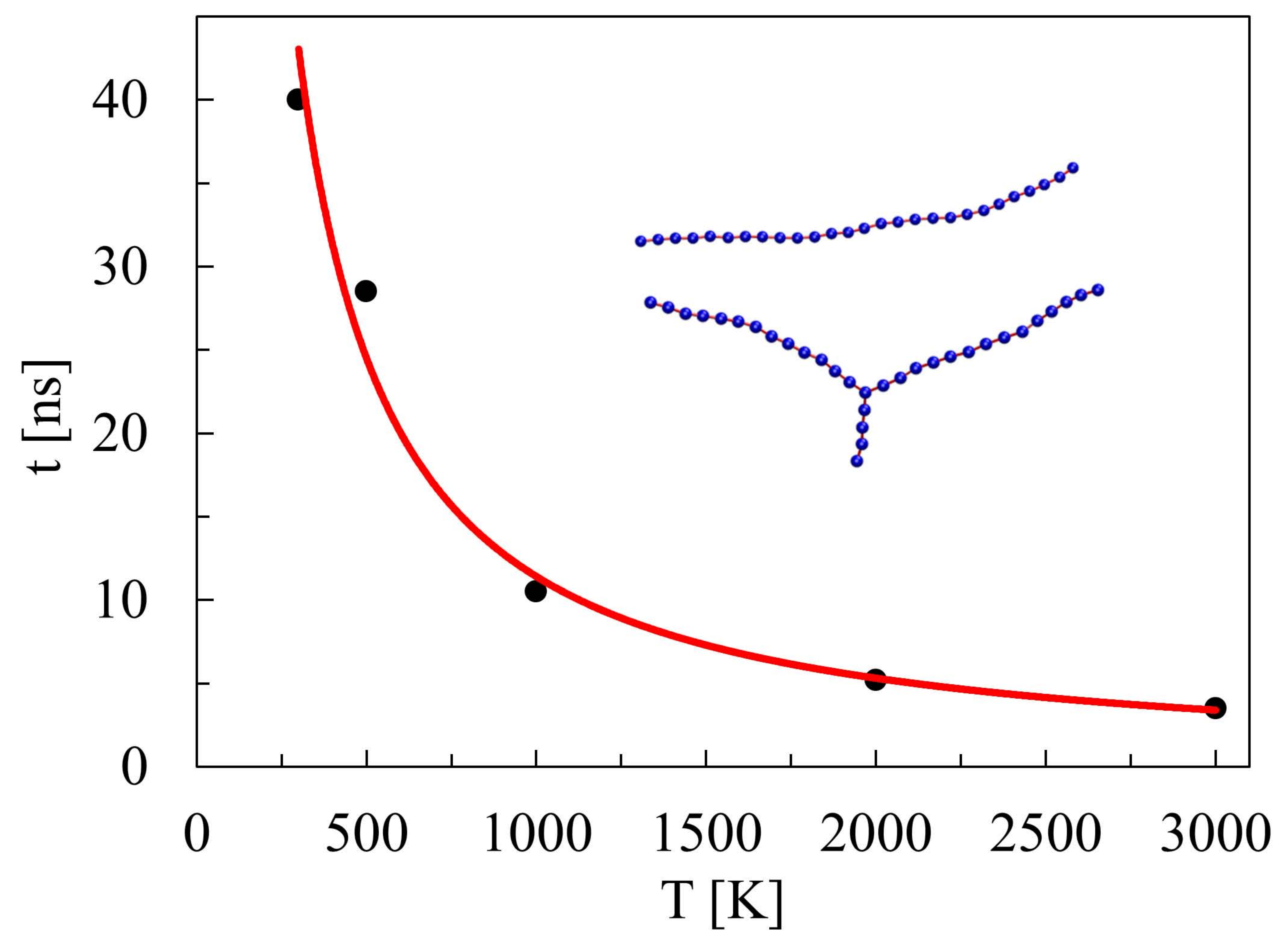} 
\caption{Nucleation time of of small
linear or branched carbon chains (inset: linear and branched carbon chains consist 
of $n_C \approx 30$ atoms) at different 
temperatures. The analytic dependence is $t(ns)= 23,153  T^{-1.10}$(K). }
\label{chain} \end{figure}

While formation of structure at much lower temperatures and densities is
hampered by slow diffusion and collision rates, it is still possible to
glean valuable information under these conditions by investigating the
pattern at higher temperatures and densities. We calculated the time
taken to form small linear or branched chain, with a carbon atom number,
($n_C \approx 30$), at different temperatures. As shown in
Fig.~\ref{chain}, at $T=300$ K, small chains form after $t \approx 40$
ns, whereas at $T= 3,000$ K, it took $t \approx 3.5$ ns to form. The
formation time at $500$, $1,000$, $2,000$ K, are $28.5$, $10.5$, $5.2$
ns, respectively. The analytic dependence between the time (t) and the
temperature (T) is $t$(ns)= $23,153$ T$^{-1.10}$(K).


Next, we measured the time of formation of small graphene flakes at
different temperatures. We observed graphene flakes after $ \approx 9.7$
ns at $T=3,000$ K, whereas at $T=1,750$ K, we noticed formation of graphene
flakes after $t=18.4$ ns. We did not observed graphene flakes forming at $T
< 1,500$ K. This is likely due to the existence of a barrier below $T <
1,750$ K.  It is then possible to obtain a fit to the formation time vs.
temperature. The analytical form is $t(ns)$=$339,404$ T$^{-1.31}$(K).
This fit predicts that a graphene flake, assuming that the classical over
the barrier can be penetrated, would form after $t(ns)$ $\sim 193$ at
$T=300$ K. 

Next, we study the effect of particle (carbon atom) density on the
structure of carbon cages/clusters.  In order to test the effect of
particle density on the structure of carbon clusters, $512$ carbons atoms
were placed in the gas phase in a cubic box with periodic boundary
conditions, where the density of carbon atoms was changed ($4.1 \times
10^{-10}$ -- $5.8 \times 10^{-7}$ \AA$^{-3}$) by varying the box size at
$T = 3,000$ K. When the concentration is $5.8 \times 10^{-10}$
\AA$^{-3}$, we find only very short carbon chain molecules ($n< 10$)
after $t=450$ ns; no planar and fullerene type clusters are observed.
However, when we increase the concentration of carbon atoms ($\rho = 5.1
\times 10^{-7}$ \AA$^{-3}$), short chain molecules emerge after $t=100$ ns.
After $t=450$ ns, long chain molecules with few cage like clusters are
observed. Fullerene type clusters are observed after $1 \mu s$.
Interestingly, no graphene flakes formed at $T = 3,000$ K. Graphene
flakes formation are observed after $t=97$ ns and $t=125$ ns at $T=2,500$ K and
$T=2,000$ K, respectively. No graphene flakes are observed at $T=300$ K and
$T=500$ K after $ t \approx 3\, \mu s$. {It should be mentioned that 
while our lowest density only extends down to $\approx 10^{14}$ cm$^{-3}$, 
still many orders of magnitude more dense than typical interstellar cloud 
densities of $10^4$ cm$^{-3}$, our simulations and scaling dependencies 
point the way to arbitrarily low temperature and density conditions.}

\subsection{Conditions at chemical equilibrium}
The Gibbs free energy of any system can be written as: 
\begin{equation} G = H - TS \,, \ \ \    S = \frac{U}{T} + k\ {\rm ln}\, Q
\, , \label{free1} \end{equation}
where $H$ and $S$ are its enthalpy and entropy, $U$ and $Q$ are its
internal energy and a canonical partition function, and $k$ is the
Boltzmann constant, respectively. Using a molecular partition function,
$q_{tot}$, $Q$ can be written as
\begin{eqnarray} Q = \frac{q_{tot}^N}{N!}\, , \ \ & q_{tot} = q_{t}\
q_{r}\ q_{v}\ q_{e}  \, \nonumber  \\ 
\label{free2} \end{eqnarray}
Here, $N$ is the number of particles, $q_{t}$, $q_{r}$, $q_{v}$, and
$q_{e}$ are the translational, rotational, vibrational and electronic
partition function. Using the partition functions, one can calculate the
free energy of a gas at a a particular temperature and
volume. The competition between the entropy and enthalpy of the system
determines the formation probability of a particular cluster as the 
system approaches chemical equilibrium. 

 Under equilibrium conditions, the free
energy of one mole of $C_2$ molecules ($G_{C_2}$) is equal to the free
energy of two moles of isolated carbon atoms ($G_{2C}$), $G_{2C} =
G_{C_2}$.  
The symmetry number ($\sigma$) for a diatomic molecule and
the $C-C$ bond distance are 2 and $1.54$ {\AA}, respectively. The bond
energy ($H_{C-C}$) and the vibrational frequency (${\nu}$) of $C-C$ are
$\epsilon_D = 348$ kJ/mol (29090 $cm^{-1}$) and $1200$ cm$^{-1}$, respectively. 

The equilibrium constant for each mol of the reaction $C_2 \rightarrow 2C$ 
is defined as the ratio of partition functions, $k_p = \frac{(q_C/N_A)^2}
{(q_{C_2}/N_A)} e^{-\beta \epsilon_D}$, where the partition functions 
$q_C = q_t$, $q_{C_2} = q_t\ q_r\ q_v$ and $N_A$ is the Avogadro's 
number \cite{laidler03}. For the C-C bond energy of $\epsilon_D$, and ideal 
gas conditions, one obtains a constant $k_p=5.4\times 10^{-33}$ at one atm 
of pressure and $T=500$ K, and a volume of dissociation of $2.6\times 10^6$ 
m$^3$ per molecule. The immediate implication is that at equilibrium,  
only high entropy structures, such as filaments and small clusters survive 
at low densities.

\vspace{-2mm} 
\subsection{Size distributions of carbon clusters}
\vspace{-2mm} 
The carbon structures which form in simulations can have varying distributions
according to atom size, moiety, ring size (pentagon, hexagon) and bond length,
depending on the density and temperature of the gas. There can be large
variations in different clusters which form. We have determined the probability
distributions on structure size (C atom number) at different temperatures and
densities. To this end, we simulated two systems with $4,100$ and $10,000$
carbon atoms (carbon density, $\rho =4.0 \times 10^{-6}$ {\AA}$^{-3}$ same for
both systems) in gas phase at $T = 3,000$, $2,000$, $1,000$, and $500$ K.
At $T= 500$ K, the size distribution is obtained after $t=200$ ns. Similarly, for
$T=1,000$, $2,000$, and $3,000$ K, the size distributions are obtained after
$t=100$, $50$, and $40$ ns, respectively. For a particular system, the
total number of formed clusters are divided into bins with each bin containing
$25$ carbon atoms. Clusters, having less than five carbons atoms are ignored.
The different statistical distributions are modeled with Gamma distribution
functions \cite{mathworld} and the parameters for each distribution are given
in the figure caption. { A Gamma distribution in a two-parameter family 
of continuous statistical distribution, related to Poisson distributed 
events which are correlated. The probability distribution function can be 
written as $P(x) = \frac{(x^\alpha -1) \exp{(-x/\alpha)}}{\Gamma{(\alpha)}
\theta^\alpha}$, with $\alpha$ and $\theta$ as the shape and scale parameters, 
and $\Gamma (\alpha)$ the usual Gamma function.}

\begin{figure} 
\vspace{2mm} 
\plotone{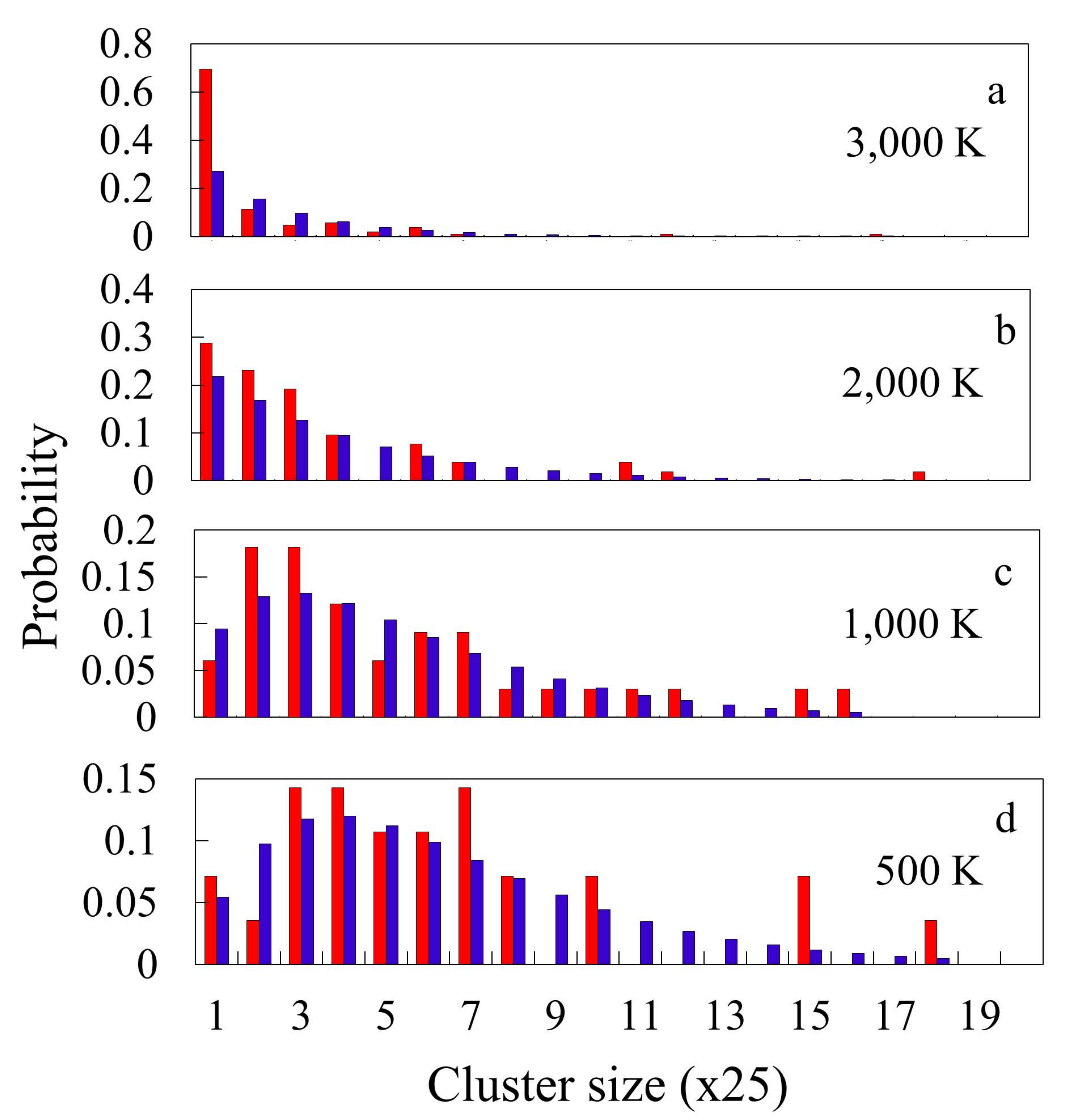} 
\caption{ Probability distribution of
carbon clusters (based on atom numbers in each cluster).  (a) Simulated and
gamma (blue; $\alpha=0.7571$ and $\theta=2.5915$) probability distributions of
obtained cluster sizes, when $4,100$ carbon atoms (atom density, $\rho =4.0
\times 10^{-6}$ {\AA}$^{-3}$) are simulated at $T = 3,000$ K. Snapshot taken
at $t=40$ ns.  (b) Simulated and gamma (blue; $\alpha=1.0832$ and
$\theta=3.177804$) probability distributions of obtained cluster sizes, when
$4,100$ carbon atoms (atom density, $\rho =4.0 \times 10^{-6}$ {\AA}$^{-3}$)
are simulated at $T= 2,000$ K. Snapshot taken at $t=50$ ns. (c) Simulated and
gamma (blue; $\alpha=2.0007$ and $\theta=2.6959$) probability distributions of
formed cluster sizes with $4,100$ carbon atoms (atom density, $\rho =4.0
\times 10^{-6}$ {\AA}$^{-3}$) at $T= 1,000$ K.  Snapshot taken at $t=100$ ns.
(d) Simulated and gamma (blue; $\alpha=2.3902$ and  $\theta=2.6447$)
probability distributions of formed cluster sizes with $4,100$ carbon atoms
(atom density, $\rho =4.0 \times 10^{-6}$ {\AA}$^{-3}$) at $T=500$ K. Snapshot
taken at $t=200$ ns. Each bin contains $25$ carbon atoms.}   
\label{dis-all4}
\end{figure}

Fig.~\ref{dis-all4}(a), shows the probability distribution of carbon clusters at
$T=3,000$ K ($4,100$ C atoms and simulation time of $t=40$ ns). We find
that carbon atoms formed mainly small clusters along with few curved and
fullerene type clusters. At high temperatures, large clusters are destabilized
by the high thermal energy. Here, the entropy dominates over enthalpy in the 
Gibbs free energy of the system. The high formation probability
of the very small clusters ($n_C< 20$) is confirmed by the probability
distribution analysis in Fig.~\ref{dis-all4}(a). At $T=2,000$ K,
although carbon atoms form mainly small clusters ($n_C\approx 35$), 
small graphene flakes and few large curved clusters, with five- and
six-member rings (bowl shape) are observed along with chains molecules (as
shown in Fig.~\ref{dis-all4}(b) with $4,100$ C atoms, sampled at $t=50$ ns of
simulations). At $T=3,000$ K and $T=2,000$ K, the carbon atoms are still in
gaseous phase (in which small clusters were mainly present) after $40$ and $50$
ns, respectively.  It is also important to note that high temperatures are
required to overcome the barrier to form curved structures to flat structures.

Next, we determine the probability distribution for $4,100$ C atoms,
(Fig.~\ref{dis-all4}(c) and $10,000$ C atoms (data not shown/see supporting
information) at $T=1,000$ K, sampled at $t=100$ ns. In this case, we observe that
the most probable carbon clusters consist of $n_C\approx 75$ carbon atoms. As
shown in Fig.~\ref{dis-all4}(d) with $4,100$ atoms (for $10,000$ C atoms data 
not shown/see supporting information) at $T=500$ K sampled at $t=200$ ns, 
carbon clusters containing $ n_C\approx 100$ carbon atoms more likely form. 
For $10,000$ atoms at $T=500$ K, sampled at $t=200$ ns, the same probability 
distribution of carbon clusters is found. The maxima of the probability 
distribution also indicates the emergence of a ``liquid phase" (in which 
relatively large clusters are present) at $T=500$ K after $t=200$ ns. For 
longer equilibration times, several small clusters can merge to big clusters 
at low temperatures.  

{At low temperatures, carbon atoms initially form short chains and then several chains coalesce to form long chain molecules. Large cluster type molecules with large member carbon rings emerge from several long chains. Eventually, these large clusters reorganize and form graphene type sheets or bent clusters. At high temperatures, initially, short chain molecules/fragments are formed from carbon atoms which then condense to form small clusters with relatively small member carbon rings. Eventually, these small clusters merge and form graphene or fullerene type molecules.}

\vspace{-4mm} \subsection{Hydrogenation process} 
\begin{figure} 
\vspace{2mm} 
\plotone{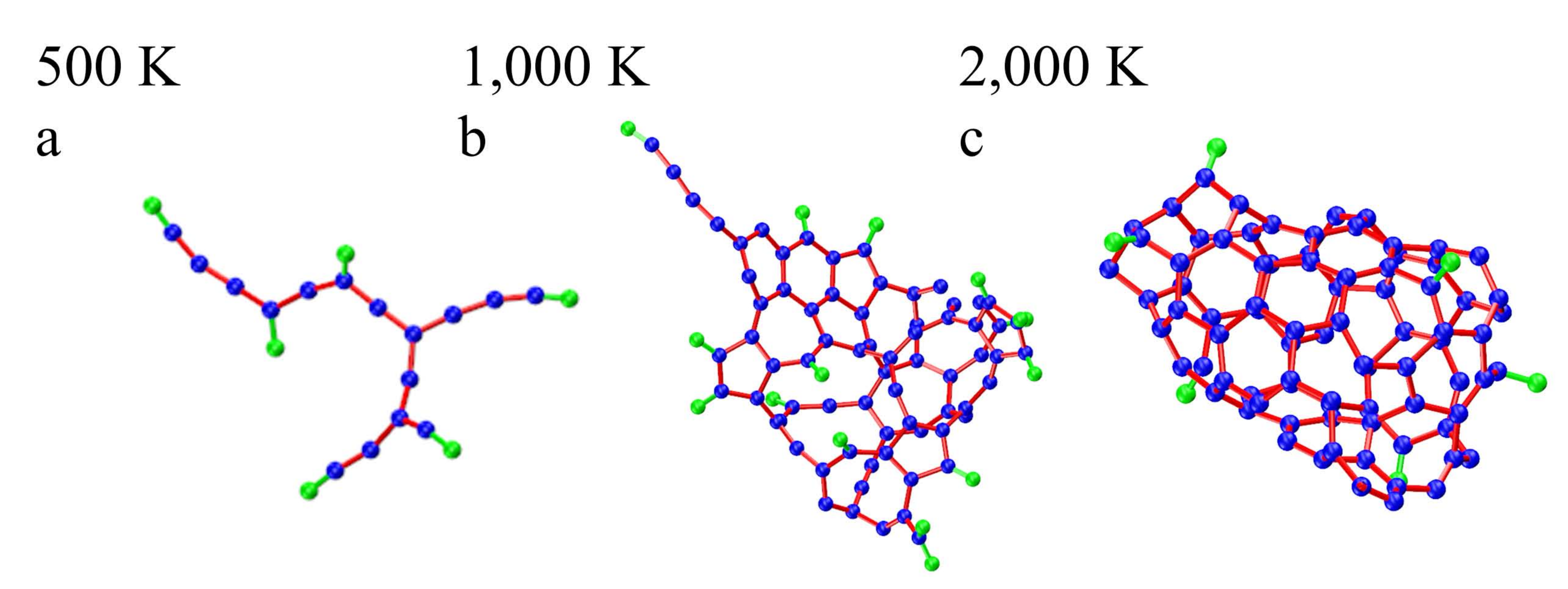} 
\caption{Formation of
Carbon-hydrogen clusters (most dominant structures are shown, snapshots
taken at $t=200$ ns). 256 carbon atoms and 256 hydrogen atoms were placed in
cubic box ($500 \times 500 \times 500$ \AA $^3$).  a) Short unsaturated
chain and branched molecules, terminated with hydrogen atoms, are formed
at $T = 500$ K. b) Planar clusters, terminated with hydrogen atoms, are
observed at $T = 1,000$ K. c) At $T = 2,000$ K, fullerene like clusters,
with some hydrogen atoms attached to surface, are found. }
\label{C256-H256} \end{figure}

\vspace{-2mm} 
In order to investigate the influence of hydrogen atom addition on the
structure of carbon cages/clusters, we simulate two systems where the ratio
between carbon atoms and hydrogen atoms are $1:1$ and $1:2$, respectively.  We
also investigate the effect of concentration on the self-assembled structures
by varying the atomic concentration. {This aspect of the study is not 
directly relevant to ISM conditions, as the C:H ratio in the ISM is about 
10$^{-4}$. The following simulations illustrate how the cluster growth 
terminates when hydrogen atoms are introduced.} For each study, after a short
minimization, we simulate the cluster formation at different temperatures.
Here, we notice the formation of alkene and other unsaturated carbon-hydrogen
chain molecules along with small carbon clusters and hydrogen molecules. We
observe that hydrogen atoms terminate the growth of big carbon clusters.  At
low temperature $T = 500$ K, small chains (terminated with hydrogen atoms) are
predominantly found (Fig.~\ref{C256-H256}(a)). Planar clusters
(Fig.~\ref{C256-H256}(b)), terminated by hydrogen atoms, are observed at $T =
1,000$ K. Fullerene type clusters are observed at $T=2,000$ K
(few hydrogen atoms are bonded on the surface), as shown in
Fig.~\ref{C256-H256}(c). {We also observed vinyl type polymers (not shown), in addition to linear alkyne chains.} At low temperatures, both the C-H and C-C bonds
formation are favorable, but at high temperatures C-H bonds formation is less
favorable compared to C-C bonds due to high thermal fluctuation.  Therefore,
small hydrogen terminated carbon chains are observed at low temperatures and
graphene to fullerene type clusters were formed at high temperatures.


\vspace{-6mm} 
\subsection{Carbon inflow: non-equilibrium nucleation}
\vspace{-2mm} 

While in a chemical steady-state condition, significant curvature of 
graphitic sheets leading to spontaneous folding, may not occur at certain 
low temperatures, inflow of carbon material could facilitate approach 
to additional equilibria to curved carbon clusters and finally to 
fullerene formation. This situation is more resembling of astrophysical 
environments. {The process of carbon in-flow in the ISM may not 
necessarily follow a Boltzmann distribution. We intend to understand 
how the carbon in-flow influences the formation process.} To this end, 
we simulate with $100$ gaseous carbon atoms 
in a $150$ {\AA$^3$} box (atom density, $\rho = 2.96 \times 10$$^{-5}$ 
\AA$^{-3}$) at $T=1,750$ K (all other conditions are the same as describe 
above). After equilibrating for $t=10$ ns, we save the coordinates of 
all atoms at the last frame and add another $20$ carbon atoms, keeping 
the box size fixed (total carbon atoms = $120$; atom density, $\rho = 
3.55 \times 10$$^{-5}$ \AA$^{-3}$). After minimization, we simulate 
for another $t=10$ ns equilibrating at $T=1,750$ K and save the coordinates 
of all atoms of the last frame and repeat the process one more time 
(total carbon atoms = $140$; atom density, $\rho = 4.15 \times 10$$^{-5}$ 
\AA$^{-3}$).   

\begin{figure} 
\plotone{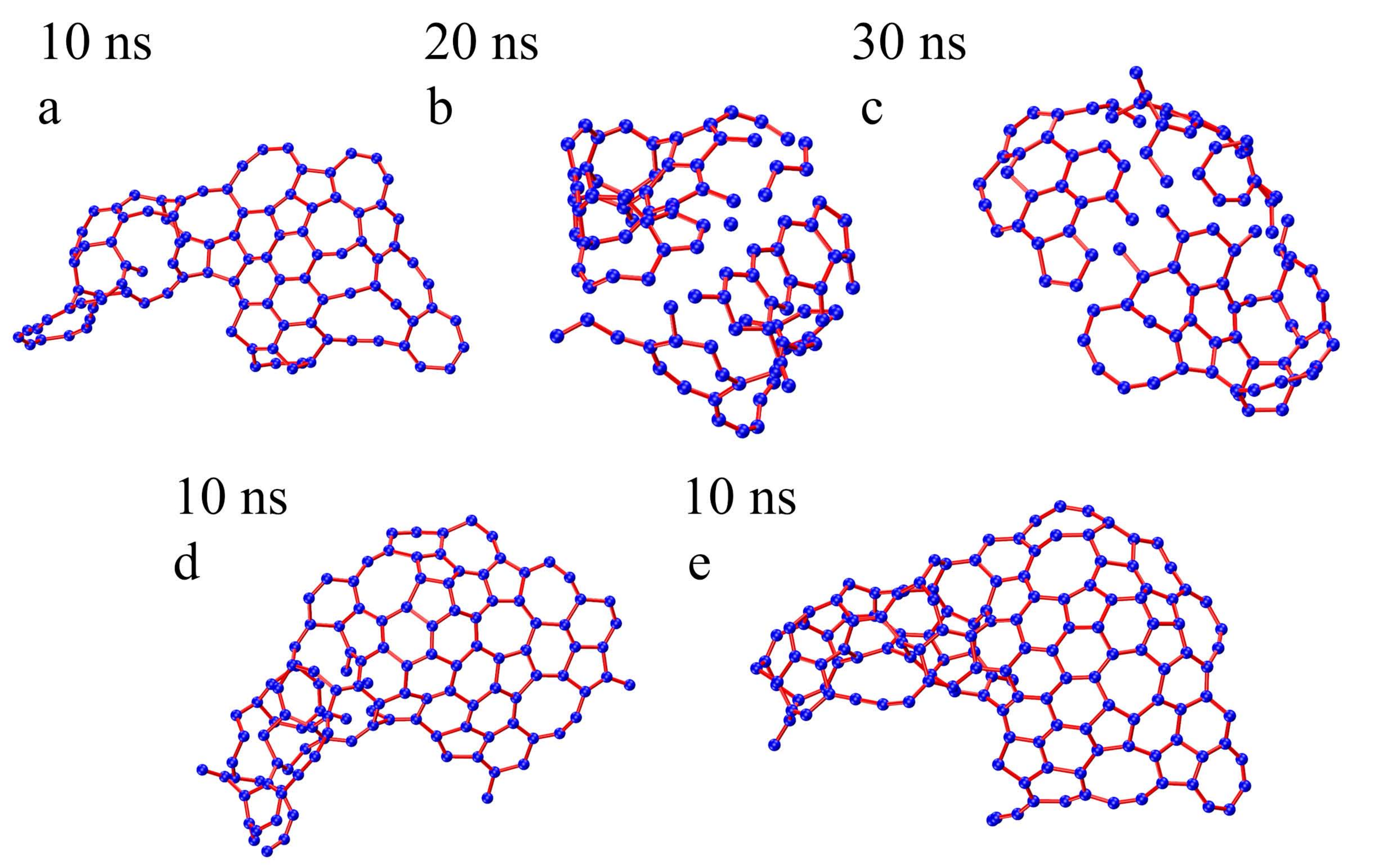}
\caption{Carbon inflow at $T = 1,750$
K. $100$ carbon atoms were placed in a cubic box ($150 \times 150 \times
150$ \AA$^3$).  a) Formation of planar cluster with few six and five
member rings is observed; snapshot taken at $t=10$ ns. b) Bend cluster, with
of six and five member rings, is observed after $t=20$ ns. c) More bend
structure, with all five and six member rings, is formed after $t=30$ ns
(note that some C-C bonds are missing for b and c; this is due to the
fact, some part of these structures are in other periodic boxes and VMD
(visualization software) can not draw bonds in such situations).
d)Formation of planar cluster after $t=10$ ns, where $20$ carbon atoms are
added after $t=10$ ns simulation of $100$ carbon atoms(a). e)Formation of
even bigger planar cluster after $t=10$ ns, where another $20$ carbon atoms
are added after $t=10$ ns simulation of $120$ carbon atoms(d).  }
\label{seed} 
\end{figure}

As shown in Fig.~\ref{seed}(a), we observe that planar cluster forms, with
few six and five member rings, after $t=10$ ns. Within $t=20$ ns, formation of
bend structure is observed with more six and five member rings
(Fig.~\ref{seed}(b)) and after $t=30$ ns, curved structures are observed where
almost all carbon atoms form six or five member rings (Fig.~\ref{seed}(c)).
When we add $20$ carbon atoms to structure in Fig.~\ref{seed}(a) and
simulate for another $t=10$ ns, we observe more planar structure compared
to Fig.~\ref{seed}(b), as shown in Fig.~\ref{seed}(d). The added carbon atoms
form bonds with edge carbon atoms of Fig.~\ref{seed}(a) and tend to create
six member rings. When we add another $20$ carbon atoms to
Fig.~\ref{seed}(d), we observe more planar structure (Fig.~\ref{seed}(e))
compared to Fig.~\ref{seed}(c), as the added carbon atoms again form bonds
with edge carbon atoms of Fig.~\ref{seed}(d) and create more six member
rings.   

\vspace{-2mm} 
\section{Conclusion}
\vspace{-2mm} 

We study conditions under which large carbon structures form and stabilize 
in the gas phase in interstellar space by using reactive MD simulations. 
The formation mechanism of the large clusters and graphene type structures 
are investigated by a series of MD simulations. The influence of the
temperature and particle density on the shape of structures are studied.
It is found that at high temperature ($T=2,000-3,000$ K), fullerene type
clusters are obtained, whereas at relatively low temperatures, graphene type
sheets along with long chain molecules/structures are synthesized. 
Analytical expressions for the small chain molecules formation can be 
used to predict the time scale to formation of nanostructures
in interstellar medium.  We determine the probability distributions of
cluster sizes at different temperatures and initial C atom number and
compare them with the gamma-distribution. The effect of hydrogen atom
addition on the structure of carbon clusters is also studied. At high
temperature, fullerene type clusters are found whereas short and branched
chains molecules, terminated with hydrogen atoms, are observed at low
temperature.  At higher density of hydrogen atoms, we find that small
unsaturated carbon chains are formed. When the concentration of hydrogen
are relative low graphene type sheet (terminated by hydrogen atoms) and
small cluster are observed. Finally, we simulate the self-assembly of
carbon structures, in non-equilibrium conditions of periodic carbon
inflow, resembling the astrophysical conditions. 

These studies help 
in better understanding the tenuous gaseous conditions in the ISM for 
synthesis of large carbon structures. We find that the system entropy 
plays a significant role at extreme conditions of low temperature and low 
density in hindering synthesis and stabilization of large carbon structures.
In future studies, we intend to investigate how dynamical quantum tunneling and 
stellar irradiation can lead to new pathways of formation of carbon clusters. {UV irradiation and shock waves can ionize the gas and play significantly in molecular synthesis in the ISM. Quantum chemical calculations are necessary to address and understand the ionization process. When the quantum chemical calculations are limited by the system size, classical (semi-empirical) methods with improved force field can be used to understand the nucleation process.}

\acknowledgments
N.P. acknowledges the support from the SAO Fellowship. The authors
are grateful for allocation of computer time on the Kraken Cluster 
at the NSF-XSEDE and Odyssey Clusters at Harvard University where 
parts of the simulations were conducted. Financial support was 
provided by a Smithsonian Grand Challenges Award.

\vspace{2mm}
%
The probability distribuions of cluster sizes for the system with 
$10,000$ carbon atoms at $T=10,000$ K and $T=500$ K are provied 
in supplementary information. 



\clearpage

\clearpage

\clearpage

\clearpage

\clearpage

\end{document}